# Structural heterogeneity-induced enhancement of transverse magneto-thermoelectric conversion revealed by thermoelectric imaging in functionally graded materials

Sang J. Park[1,*], Ravi Gautam[1], Takashi Yagi[2], Rajkumar Modak[3], Hossein Sepehri-Amin[1] and Ken-ichi Uchida[1,3,*]

[1] National Institute for Materials Science, Tsukuba 305-0047, Japan

[2] National Institute of Advanced Industrial Science and Technology, Tsukuba 305-8563, Japan

[3] Department of Advanced Materials Science, Graduate School of Frontier Sciences, The University of Tokyo, Kashiwa 277-8561, Japan

*Correspondence to: PARK.SangJun@nims.go.jp (S.J.P.);
UCHIDA.Kenichi@nims.go.jp (K.U.)

**Abstract**

Functionally graded materials (FGMs) exhibit continuous property variations that enable unique functionalities and provide efficient platforms for systematic property optimization. Here, we report the fabrication of FGMs with graded structural heterogeneity by annealing an amorphous metal under a one-dimensional temperature gradient. Using lock-in thermography (LIT), we spatially mapped transverse thermoelectric conversion with high spatial and temperature resolution. A pronounced non-monotonic response was observed, with the maximum anomalous Ettingshausen effect, transverse charge-to-heat conversion in magnetic materials, appearing in the atomic-heterogeneity regime well before crystallization. This enhancement was not captured by conventional structural or longitudinal transport measurements, highlighting the exceptional sensitivity of transverse thermoelectric phenomena to subtle structural heterogeneity. Structural analyses using scanning transmission electron microscopy and atom probe tomography revealed Fe-based crystalline alloys and Cu nanoclusters embedded in the amorphous matrix, whose heterogeneity accounts for the enhanced response. These findings establish temperature-gradient-annealed FGMs, combined with LIT, as a powerful methodology for probing structural-heterogeneity-driven transverse electron transport and designing high-performance flexible materials.

1. **Introduction**

Functionally graded materials (FGMs) are a class of advanced materials characterized by continuous gradients in composition, microstructure, or porosity along spatial directions [1–5]. Such continuous variations in material properties impart unique functionalities, such as enhanced mechanical performance [5,6] and efficient energy conversion within solids [7–9], by spatially optimizing material properties. FGMs have been widely adopted across diverse fields, including high-performance composites for aerospace, marine, and nuclear applications [1,10], and energy materials for efficient thermoelectric conversion [7–9]. They offer potential solutions to the limitations of conventional homogeneous materials and heterostructures by overcoming intrinsic property constraints.

Moreover, FGMs provide a unique experimental platform for investigating material properties, owing to their continuous property gradients. If these gradients can be spatially resolved with sufficient precision, efficient parameter studies can be performed by treating the property variations as experimental variables within a single sample. This approach enables characterization and optimization with significantly improved cost-effectiveness and throughput compared to conventional strategies, which typically rely on discrete sample sets and limited parameter sweeps [11–15], thereby often underestimating the maximum potential of materials.

A promising technique for characterizing such continuously varying properties is contactless transport measurement using lock-in thermography (LIT). LIT captures the thermal response of a solid to periodic excitation, offering high temperature sensitivity (<1 mK) [16] and effective separation of the periodic thermal signals from background temperature fluctuations [17], compared to conventional steady-state infrared (IR) measurements [18–20].

These capabilities enable site-specific detection of microscale defects and dislocations in materials such as graphene [17], as well as intrinsic thermoelectric properties such as Seebeck and Peltier coefficients [21,22]. Recently, LIT has also been used to study transverse thermoelectric effects [23–28], such as the anomalous Ettingshausen effects (AEE), which represents the transverse generation of temperature gradient in magnetic materials in response to an applied longitudinal charge current [29–31].

There have been studies aimed at optimizing the transverse thermoelectric properties by integrating the FGMs with LIT. For instance, Modak et al. [25] fabricated a Heusler $Co_2MnAl_{1-d}Si_d$ combinatorial film, where the relative compositions of Al (1-$d$) and Si ($d$) were gradually varied along one direction. This FGM film provided a novel platform to investigate the role of chemical composition and Fermi level control in AEE. Using LIT measurements, the optimal AEE performance was revealed specifically in the range $0.06 < d < 0.12$, which was higher than that of constituent homogeneous $Co_2MnAl$ ($d = 0$) and $Co_2MnSi$ ($d = 1$) films. Similar approaches have been successfully applied to the development of high-performance spin Hall materials using $Cu_dIr_{100-d}$ [32] and $Pt_dW_{100-d}$ [24] alloy thin films as well as magnetic $Sm_dCo_{100-d}$ and $Sm_{20}(Co_{100-d}Fe_d)_{80}$ thin films [33], all of which benefited from the integration of FGMs and LIT. The key ideas of these demonstrations are (1) the precise control of transverse deflection mechanisms of electrons in solids [34–39] as experimental variables, such as Berry-curvature-driven intrinsic mechanisms and/or extrinsic mechanisms (e.g., skew scattering and side jump mechanisms) within a single FGM sample, and (2) the high-resolution spatial detection of their effects on transverse thermoelectric conversion by measuring the AEE with spatial resolution, enabled by LIT.

Recently, large transverse thermoelectric conversion has been observed in structurally heterogeneous magnetic composites [40–45], with magnitudes comparable to those in single-crystalline topological materials or in epitaxial thin films grown on rigid platforms [30,31,46–53]. Interestingly, such large conversion was demonstrated while retaining mechanical flexibility, suggesting potential applications based on the anomalous Nernst effect (ANE), which is the reciprocal of the AEE and corresponds to transverse heat-to-charge current conversion in magnetic materials. Applications include thermal energy harvesting [40–43,54] and heat flux sensing [44,55–58] from various curved heat sources, such as heat pipes. A recent study combining theoretical model and experimental validation suggested that the large transverse thermoelectric conversion originates from the formation of heterogeneous multi-phase structures [41], which cannot be explained by the above-mentioned conventional theoretical intrinsic and extrinsic frameworks [34–39]. Independent experimental studies have also shown that the heterogeneous formation of nanoclusters and/or local structural disorder in an amorphous matrix effectively increases the transverse thermoelectric conversion [40,42,43]. All these results collectively suggest the important role of structural engineering in transverse thermoelectric conversion, while the detailed mechanisms based on a specific formulation are yet to be fully understood in the complex system.

If the structural characteristic can be spatially controlled within a single sample by fabricating an FGM and the corresponding variation in transverse thermoelectric response is mapped using LIT, the effects of structural heterogeneity on AEE and ANE can be investigated systematically in a statistically significant and high-throughput manner. In addition, such FGMs serve as an ideal platform for determining the optimal degree of heterogeneity that maximizes AEE and ANE, thereby providing high-resolution insight into the relationship between structural heterogeneity and transverse electron transport mechanisms.

Here, we report the observation of transverse thermoelectric conversion in structural-heterogeneity-graded FGM ribbons by measuring the spatial distribution of AEE using LIT. First, we introduce a simple and scalable fabrication method for bulk FGM ribbons by annealing an Fe-based fully amorphous ribbon under a temperature gradient (**Fig. 1**), where annealing-temperature-driven structural heterogeneity can tune the transverse thermoelectric property [40–43]. We then demonstrate a continuous variation of the AEE response with annealing temperature in the single sample, exhibiting a non-monotonic trend in which the AEE reaches a maximum at a mid annealing temperature well before crystallization.

Through structural analyses using atom probe tomography (APT) and scanning transmission electron microscopy (STEM), we clarified that the enhanced AEE is strongly correlated with the presence of Fe-based crystalline alloys and/or Cu nanoclusters embedded in the amorphous matrix, providing compelling evidence for the critical role of structural heterogeneity in enhancing transverse electron scattering. This correlation offers design guidelines for transverse thermoelectric materials based on magnetic amorphous alloys. Notably, such subtle structural variations responsible for the AEE enhancement are not captured by conventional characterization techniques such as lab-scale X-ray diffraction (XRD) or longitudinal electrical and thermal transport property measurements, highlighting the unique sensitivity of transverse thermoelectric phenomena to atomic and nanoscale structural heterogeneity.

## 2. Results and discussion

### 2.1 Fabrication of heterogeneity-graded FGM

We first outline the design strategy for fabricating bulk heterogeneity-graded FGMs using amorphous ribbons. An Fe-based amorphous alloy with the composition $Fe_{79}Si_4B_{14}Nb_2Cu_1$ was selected as the demonstration material, as it has been reported to exhibit large AEE under controlled annealing temperatures ($T_a$) while maintaining mechanical flexibility [40,42]. Specifically, the AEE peaks when the sample is annealed in the $T_a$ range of 523–673 K, and decreases significantly at higher $T_a$, indicating the importance of annealing-induced structural heterogeneity, including crystallinity, precipitation, and atomic-scale disorder, for optimizing AEE [42]. The broad $T_a$ window associated with enhanced AEE performance (523–673 K) also allows for a clear investigation of the relationship between structural change and AEE, making this alloy system a suitable platform for demonstration.

The heterogeneity-graded FGM was fabricated by annealing a long bulk amorphous ribbon under a temperature gradient naturally formed inside a horizontal furnace (see setup photograph in **Fig. S1**). The initial amorphous ribbon was prepared by the melt-spinning method and cut into ~130 mm in length (**Methods**). The furnace consisted of two temperature zones: a uniform heating zone controlled by PID-controlled heating element and a cooling zone maintained by a fan, which also protected the O-ring joint used for vacuum sealing. When the heating zone was set to a high temperature, a temperature gradient developed between the two zones via thermal conduction. The annealing was performed under high vacuum ($<10^{-4}$ Pa) to ensure a one-dimensional uniform temperature gradient along the ribbon length and to prevent oxidation. One end of the ribbon was placed in the heating zone and the other in the cooling zone, resulting in a continuous gradient of structural heterogeneity along its length.

The temperature range for annealing was determined based on differential scanning calorimetry (DSC). The DSC data in **Fig. S2** show two exothermic peaks during heating at approximately 711 K and 847 K, corresponding to crystallization events. Based on this, the heating zone temperature was set to 773 K, which produces a broad range of structural states from the amorphous on cooling-zone side to partially crystalline on heating-zone side, as illustrated in **Fig. 1**. To enhance cooling and maximize the temperature gradient, the exterior of the cooling zone was covered with a wet cloth (photograph in **Fig. S1b**). Additional details on the fabrication procedure and experimental setup are provided in **Methods**. It is worth noting that although the temperature gradient may not be strictly linear, it remained continuous, producing a continuous gradient of structural states. In the following sections, we examine how this annealing under a temperature gradient influences crystal structures and various material properties, including both longitudinal and transverse transport properties.

## 2.2 Characterization of structural heterogeneity and longitudinal transport properties

We investigated the gradual change in structural heterogeneity using XRD measurements at multiple positions along the sample. The 130 mm-long heterogeneity-graded FGM ribbon was cut to fit the XRD holder. XRD measurements were performed at 10 mm intervals along the sample length to examine the correlation between $T_a$ and crystallinity. The $x$-position in **Fig. 2** denotes the distance from the low temperature side ($T_L$) to the high temperature side ($T_H$). This positional reference is used consistently throughout all position-dependent data in this study. XRD patterns from the $T_L$ side ($x = 5$ mm) exhibit broad peaks in the $2\theta$ range from 30-75° (based on Cu-K$\alpha_1$ radiation; **Methods**), indicating a fully amorphous structure (**Fig. 2a**). These amorphous features persist up to $x = 100$ mm, beyond which sharp diffraction peaks corresponding to α-Fe appear, with negligible evidence of other phases. To quantify the crystallization onset, the full width at half maximum (FWHM) was extracted from each diffraction pattern and presented in **Fig. 2c**. A sharp decrease in FWHM begins at approximately $x =110$ mm, indicating the onset of crystallization, whereas a small variation is observed across the 0–100 mm region. Beyond this point, the sample exhibits broad amorphous features together with emerging crystalline peaks, indicating partial crystallization from fully amorphous state into amorphous and *α*-Fe nanocrystalline phases and the resulting structural heterogeneity (as illustrated in **Fig. 1**). The detailed XRD data are provided in **Fig. S3**. The structural changes induced by annealing will be discussed in greater detail using STEM and APT measurements in a later section.

We next measured the longitudinal electrical conductivity ($\sigma$) at $x = 5$ mm intervals using the same samples used for the XRD measurements, where the samples were mounted on glass plates. A standard four-probe method was employed, with two voltage-sensing needle probes scanned along the samples in 5 mm steps, while current electrodes were fixed at both ends (inset, **Fig. 2d**). Amorphous phases are known to exhibit lower $\sigma$ than crystalline phases due to electron localization, and $\sigma$ increases upon crystallization [40,41]. **Figure 2d** shows that $\sigma$ remains nearly constant at ~7.6 × $10^5$ S/m from $x$ = 0–110 mm and then increases significantly beyond $x$ = 110 mm, reaching 14.1 × $10^5$ S/m at $x$ = 125 mm, consistent with the structural changes observed by XRD.

We then measured the longitudinal thermal diffusivity ($D$). Similar to $\sigma$, $D$ is expected to be low in the amorphous regime and higher in the crystalline regime due to enhanced heat transport from both lattice and electrons [40–43]. Position-dependent $D$ was measured using the spot periodic heating radiation thermometry method [59]. In this method, the center of a suspended sample was periodically heated by a laser beam at a frequency $f$, inducing thermal oscillations with a finite thermal diffusion length. Given the small sample thickness (20 ± 2 μm), heat diffusion through the thickness (i.e., $z$-direction) was assumed to be instantaneous, allowing the heat transfer to be approximated as radial heat flow in the sample plane. Radial heat diffusion was detected using a temperature sensor (**Methods**), which measured both the amplitude and phase lag ($\phi_{lag}$) of the temperature oscillations relative to the modulated laser input. The phase signal reflects the time delay of heat propagation through the material and was used to extract $D$ using the following equation [59]:

$$D = \pi f / (\frac{d\phi_{lag}}{dl})^2, \tag{1}$$

where $l$ is the distance from the thermal excitation center. The phase profile was measured along the $y$-direction (see inset, **Fig. 2e**), which is orthogonal to the annealing direction ($x$). Material properties are assumed to be spatially uniform due to the absence of a structural heterogeneity gradient in $y$. Further details of the measurement are provided in **Methods**.

Thermal diffusivity was measured at $x = 5$ mm intervals. **Figure 2b** shows the $\phi_{\text{lag}}$ profile of the temperature oscillation measured at $f = 5$ Hz. The slope ($\frac{d\phi_{\text{lag}}}{dl}$) was extracted from the linear regimes in the $y$-position ($\pm 1$ mm to $\pm 0.25$ mm), and $D$ was calculated using **Eq. (1)**. The symmetric and linear $\phi_{\text{lag}}$ data confirm negligible variation of thermal properties in the $y$-direction, as expected from the one-directional ($x$) temperature gradient during annealing. **Figure 2e** shows that $D$ remains constant at approximately 2.1 mm$^2$/s from $x = 0$ to 110 mm, then increases sharply beyond $x = 110$ mm, reaching 2.8 mm$^2$/s at $x = 125$ mm and 3.6 mm$^2$/s at $x = 130$ mm, consistent with increased crystallinity and enhanced lattice and electron contributions [60].

As shown above, XRD, electrical conductivity, and thermal diffusivity measurements consistently show that annealing an amorphous ferromagnetic metal ribbon under a temperature gradient leads to structural and transport property changes detected mainly at the crystallization onset near $x = 110$ mm, whereas the continuous gradient of structural heterogeneity across the ribbon cannot be resolved by these conventional characterization methods. In the next section, we show that AEE imaging via LIT reveals continuous property variation over a much broader spatial range, highlighting its superior sensitivity to subtle structural heterogeneity.

### 2.3 Spatial mapping of transverse thermoelectric response using LIT

We now present the measurement of the AEE in the heterogeneity-graded FGM sample using LIT with high spatial resolution. A 130-mm-long sample was cut into two halves to apply a magnetic field using an electromagnet, where the sample length should be smaller than the size of the pole pieces. Both ends of the sample were connected to copper wires using silver epoxy and attached to an AC current source. The measurement field-of-view (FOV) was 7.68 mm × 9.60 mm for 512 × 640 pixels (327,680 data points in one FOV), providing high-resolution imaging and statistical data points. The entire length of the sample was scanned accordingly, as illustrated in **Fig. 1**.

LIT was performed by applying a square-wave-modulated AC current to the sample under a constant magnetic field ($H$). An IR camera measured both the lock-in amplitude ($A$) and the phase ($\phi$) of the temperature modulation induced by the applied charge current ($I$). The measured signal includes contributions from multiple effects: the AEE ($\propto j_C$), Peltier effect ($\propto j_C$), and Joule heating ($\propto j_C^2$) contributions, where $j_C$ is the charge current density. To isolate the thermoelectric contributions, Joule heating was separated by using a zero-offset square-wave-modulated current, which generates a non-oscillating thermal background due to Joule heating without periodic temperature variations (**Fig. 1**). To further extract the pure AEE, the measured data were symmetrized with respect to the magnetic field polarity (±$H$). The field-odd components of the amplitude ($A_{\text{odd}}$) and phase ($\phi_{\text{odd}}$) were extracted using the following equations [24,26,27,61–63]:

$$A_{\text{odd}} = \frac{|A_{+H}\exp(-i\phi_{+H}) - A_{-H}\exp(-i\phi_{-H})|}{2}, \tag{2-1}$$

$$\phi_{\text{odd}} = -\arg\left[\frac{(A_{+H}\exp(-i\phi_{+H}) - A_{-H}\exp(-i\phi_{-H}))}{2}\right]. \tag{2-2}$$

Here, $A_{\pm H}$ and $\phi_{\pm H}$ denote the lock-in amplitude and phase under positive (+$H$) and negative (-$H$) magnetic fields, respectively. Note that the magneto-Peltier effect is symmetric with respect to $H$, whereas the AEE exhibits field-odd dependence [24,26,27,61–63]. Therefore, $A_{\text{odd}}$ and $\phi_{\text{odd}}$ selectively represent the AEE contribution.

**Figure 3** presents the spatial mapping of $A_{\text{odd}}$ and $\phi_{\text{odd}}$, measured under an applied current of $I = 500$ mA and a modulation frequency of $f = 5$ Hz (**Methods**). Magnetic fields of ±0.3 T were applied along the *y*-direction, which is orthogonal to both the sample thickness (*z*) and the current direction (*x*), thereby satisfying the symmetry condition required to detect the AEE using LIT in the *z*-direction [27,61–63]. Due to the soft magnetic properties of the sample, with typical coercivity below 0.1 mT [40,42], the entire sample was fully magnetized under the applied *H*. The constant values of $\phi_{\text{odd}}$ in **Fig. 3b** confirm that the sample was uniformly magnetized along the *y*-direction without noticeable misalignment, validating that the signals represent the pure AEE contribution.

**Figure 3a** shows that $A_{\text{odd}}$ varies significantly along the *x*-direction, reflecting the spatial gradient in structural heterogeneity. In contrast, negligible variation is observed along the *y*-direction, indicating that the material properties change predominantly along *x*. This behavior is consistent with the one-dimensional temperature gradient imposed by the horizontal furnace (**Fig. 1**), further supporting the reliability of estimating *D* from the *y*-directional $\phi_{\text{lag}}$ (**Fig. 2b**).

To clarify the *x*-directional evolution of the transverse thermoelectric response, we extracted the $A_{\text{odd}}$ values near the center region of the samples (highlighted by the horizontal white dashed lines in **Fig. 3a**) to minimize the warping artifacts. The resulting profile is shown in **Fig. 3c**. Notably, $A_{\text{odd}}$ exhibits a non-monotonic profile: it increases steadily from $x = 0$ mm (4.90 mK), peaks at $x = 70–80$ mm (6.65 mK), and then decreases gradually, reaching 2.29 mK at $x = 130$ mm. This continuous variation clearly demonstrates the existence of a structure heterogeneity gradient along the *x*-direction across a wide range of *x*. A steep drop occurs at $x = 105$ mm, corresponding to the crystallization onset of α-Fe (**Fig. 2a**). Since α-Fe is known to exhibit small transverse thermoelectric conversion [64], this structural transition explains the sharp drop. The corresponding $\phi_{\text{odd}}$ profile remains nearly constant at 180° across the sample, with small fluctuations of $\pm 2°$ (**Fig. 3d**), indicating that the sign of the AEE signal is unchanged, consistent with previous observations [40,42]. The absence of data near $x = 75$ mm in **Figs. 3c** and **3d** is due to the inability to measure AEE signals in the electrode regions used for current injection.

This non-monotonic trend clearly highlights the unique behavior of the transverse thermoelectric properties compared to the longitudinal transport properties. In the broad region of $x = 40–105$ mm, the AEE response is higher than that in the amorphous ($T_{\text{L}}$) and crystalline ($T_{\text{H}}$) regimes. Since both longitudinal electrical and thermal transport properties remain nearly constant before crystallization ($x < 110$ mm), this enhancement directly indicates the presence of additional mechanisms driving transverse electron deflection. These mechanisms are likely linked to the gradual evolution of structural heterogeneity in the heterogeneity-graded FGM sample. Previous studies have reported enhanced transverse thermoelectric conversion in Fe-based ferromagnetic amorphous alloys annealed at mid temperatures near the crystallization temperature without strong correlation with chemical composition [42]. Such enhancements have been observed experimentally in ANE and AEE [40,42,43] as well as in the anomalous Hall effect (AHE) [41], with a theoretical model describing the electron paths in complex disordered materials [41]. The proposed mechanisms include: (1) the formation of amorphous-crystalline heterostructures [41,43], (2) enhanced spin-orbit coupling (SOC) induced by Cu nanoclusters [40,42], and (3) subtle atomic-scale disorder within the amorphous matrix [42]. These mechanisms are interrelated and collectively suggest the importance of controlling structural heterogeneity at both atomic and nanoscale dimensions in transverse transport. Our observation that the AEE maximum occurs at $x$ = 70–80 mm, well before the crystallization onset ($x$ = 110 mm), suggests that the enhancement in $T_M$ is associated with the mechanisms of (2) Cu nanoclusters. This interpretation is further supported by bright field (BF)-STEM and APT analyses as discussed in the next section. Furthermore, this enhancement in $x$ = 40–105 mm could not be captured by conventional characterization techniques such as XRD or longitudinal transport measurements (shown in **Fig. 2**), demonstrating that transverse thermoelectric effects are exceptionally sensitive to subtle structural change.

## 2.4 Structural analyses based on STEM and APT

To clarify the origin of the enhanced AEE observed between $x = 40$–$105$ mm (**Fig. 3c**), we performed detailed structural analyses using BF-STEM and APT. Three representative positions were selected at $x = 5$ mm, $80$ mm, and $125$ mm, corresponding to the amorphous ($T_L$), structurally intermediate ($T_M$), and crystalline ($T_H$) regimes with distinct AEE responses.

The STEM results in **Fig. 4a** and **4c** show that the $T_L$ sample exhibits fully amorphous phase, as evidenced by the ring-type diffraction patterns in the inset. In contrast, the $T_H$ sample exhibits mixed amorphous and Fe-crystalline phases, revealing pronounced nanoscale structural heterogeneity. These observations are consistent with the position-dependent XRD results (**Figs. 2a** and **S3**). The $T_M$ sample ($x = 80$ mm) appears amorphous without a distinct contrast in structural features compared with the $T_L$ sample (**Fig. 4b**). It should be noted that when the characteristic length scale of such heterogeneity is much smaller than the STEM specimen thickness (tens of nanometers), these nonperiodic features are difficult to resolve directly by STEM.

To capture this hidden structural information, we performed APT on the $T_M$ sample to probe three-dimensional chemical distributions (**Methods**). The APT results in **Fig. 4d** revealed the formation of Cu nanoclusters with an average diameter of 1.7 nm and a number density of $2.55 \times 10^{23}$ m$^{-3}$, whereas the other constituent elements (i.e., Fe, Si, Nb, and B) remained uniformly distributed. This indicates the presence of atomic-scale heterogeneity, consistent with our previous observations from discrete sample sets [40,43]. The emergence of these Cu nanoclusters may locally break inversion symmetry at the interfaces with the ferromagnetic Fe-based host matrix, thereby inducing interfacial Rashba-like SOC and generating asymmetric skew scattering [40,42]. This additional interfacial scattering channel enhances transverse thermoelectric conversion, providing a microscopic origin for the increased AEE in the $T_M$ regime.

**2.5 Discussion on the origin of enhanced transverse thermoelectric conversion**

To elucidate the origin of enhanced transverse signal, we evaluate the anomalous Nernst conductivity ($\alpha_{\text{ANE}}$) as a fundamental transport coefficient governing transverse thermoelectricity. The anomalous Ettingshausen coefficient determined experimentally is

$$\Pi_{\text{AEE}} = \frac{\pi}{4}\frac{\kappa \Delta T_{\text{AEE}}}{j_c t}, \tag{3}$$

where $\Delta T_{\text{AEE}}$ is the magnitude of $\Delta T$ induced by AEE along the $z$-direction (i.e., $\Delta T_{\text{AEE}} = 2A_{\text{odd}}$), $j_\text{C}$ is the longitudinal charge-current density, $t$ is the sample thickness, and $\kappa$ is the thermal conductivity [40,42,43]. The anomalous Nernst conductivity is expressed as

$$\alpha_{\text{ANE}} = \sigma_{xx}(S_{\text{ANE}} + \theta_{\text{AHE}} S_{xx}), \tag{4}$$

where $S_{\text{ANE}}$ is the anomalous Nernst coefficient, $S_{xx}$ is the Seebeck coefficient, and $\theta_{\text{AHE}}$ (=$\sigma_{\text{AHE}}/\sigma_{xx}$) is the anomalous Hall angle with $\sigma_{\text{AHE}}$ being the anomalous Hall conductivity. The Onsager reciprocal relation connects the two effects as [40,42,43]

$$\Pi_{\text{AEE}} = T S_{\text{ANE}}, \tag{5}$$

where $T$ is the absolute temperature, leading to

$$\alpha_{\text{ANE}} = \sigma_{xx}(\Pi_{\text{AEE}}/T + \theta_{\text{AHE}} S_{xx}). \tag{6}$$

The first term ($\sigma_{xx}\Pi_{\text{AEE}}/T$) represents the ANE-related intrinsic transverse thermoelectric contribution, whereas the second term ($\sigma_{xx}\theta_{\text{AHE}}S_{xx}$) corresponds to the Hall-deflected Seebeck contribution. Here, the term "intrinsic" denotes a material specific transport property governed by $S_{\text{ANE}}$ or $\Pi_{\text{AEE}}$, which should be distinct from intrinsic Berry curvature vs. extrinsic (skew/side-jump) classifications of microscopic mechanisms discussed elsewhere [34–39].

We compare these two contributions in Eq. (6) by inserting the experimentally measured $\Pi_{AEE}$, $\sigma_{AHE}$ (hence $\theta_{AHE}$), and $S_{xx}$ (Methods). By assuming negligible ordinary contributions (i.e., ordinary Nernst and Ettingshausen effects), as supported by previous studies reporting minor field-dependent components in amorphous alloys [41], we evaluate $\alpha_{ANE}$ as a fundamental transport coefficient. Following the STEM and APT analyses, we select three representative positions for comparison: $T_L$, $T_M$, and $T_H$. **Figure 5** summarizes the transport parameters: $\sigma_{AHE}$ increases towards $T_H$ mainly due to enhancement of $\sigma_{xx}$, whereas $S_{xx}$ remains nearly constant at approximately -1.3 µV/K across the samples. To quantify the relative contributions, we define a dominance ratio

$$R \equiv \left|\frac{\sigma_{xx}\Pi_{AEE}/T}{\sigma_{xx}\theta_{AHE}S_{xx}}\right| = \left|\frac{\Pi_{AEE}/T}{\theta_{AHE}S_{xx}}\right|. \tag{7}$$

Across $T_L$, $T_M$, and $T_H$, we obtain $R > 25$ (i.e., the contribution of the Hall-deflected Seebeck term is < 4%, as shown in **Fig. 5c**), demonstrating that the observed enhancement is dominated by the ANE-related term $\sigma_{xx}\Pi_{AEE}/T$, while the $\theta_{AHE}S_{xx}$ channel is negligible.

**Figure 5c** shows that the resulting $\alpha_{xz}$ increases monotonically from $T_L$ (1.30 A/mK) to $T_M$ (1.83 A/mK) and peaks at $T_H$ (2.39 A/mK). This peak value exceeds the maximum value reported for discreate sample sets with similar compositions (up to 1.93 A/mK) in previous studies [40], clearly demonstrating the effectiveness of our strategy integrating FGM with LIT. The enhanced $\alpha_{ANE}$ in $T_H$ is attributable to the nanoscale amorphous-crystalline heterostructure (as observed by STEM in **Fig. 4c** and XRD in **Fig. S3**), which may induce asymmetric carrier scattering, consistent with the recently reported composite-transport model in ref. [41].

**2.6 Mechanical flexibility of heterogeneity-graded ribbon**

Another promising functionality of amorphous alloys is their mechanical flexibility, which allows transverse thermoelectric energy conversion from curved heat sources, for example using coiled structures as proposed and demonstrated in [43,56,65,66]. This versatility broadens the applicability of the materials to diverse device architectures. It is also well established that mechanical ductility is strongly dependent on the crystallinity of the system [40,41,43]. For example, Park et al. reported that the maximum bending radius of amorphous metals increases significantly when the samples are annealed at high temperatures, which was explained by the reduction of ductile amorphous fractions and the growth of rigid crystalline fractions within the composite matrix [43]. Therefore, achieving high transverse thermoelectric performance while retaining mechanical flexibility is of great importance for the development of flexible transverse thermoelectric materials.

To further examine the mechanical property of the heterogeneity-graded FGM ribbon, we attached the sample to a curved surface with a radius of 5 mm. The same samples used for the thermal diffusivity measurement were employed for this mechanical test. **Figure S4** shows that the overall sample retained its mechanical flexibility. However, the $x > 110$ mm region fractured during handling due to the brittle crystalline phases, confirmed by the XRD (**Fig. 2**) and STEM (**Fig. 4**). This result qualitatively demonstrates that the high-performing amorphous-to-nano-heterogeneity region of the FGM is applicable to devices that require mechanical flexibility.

## 3. Conclusion

We demonstrated a simple and scalable strategy to fabricate heterogeneity-graded amorphous alloy ribbons by annealing under a one-dimensional temperature gradient. This approach provides a unique experimental platform to systematically investigate the relationship between structural heterogeneity and transverse thermoelectric conversion. By employing LIT with high spatial and temperature resolution, we directly mapped the AEE along the heterogeneity gradient and revealed a pronounced non-monotonic response of AEE-induced temperature modulation under a constant electrical current. Detailed STEM and APT analyses confirmed that this enhancement originates from nanoscale structural heterogeneity of Fe-based alloys and Cu nanoclusters within the amorphous matrix. Importantly, these critical features could not be detected by conventional characterization methods such as lab-scale XRD or longitudinal transport measurements, underscoring the exceptional sensitivity of transverse thermoelectric phenomena to subtle structural variations. Furthermore, the high-performing amorphous-to-nano-heterogeneous region of the FGM ribbon retains its mechanical flexibility, supporting its applicability in flexible device architectures. Overall, this work establishes heterogeneity-graded FGMs combined with LIT imaging as a powerful, high-throughput, and highly sensitive methodology to study the role of structural heterogeneity in transverse electron transport. Beyond advancing the fundamental understanding of heterogeneity-driven thermoelectric phenomena, our results highlight a practical route for designing flexible materials and devices with enhanced transverse thermoelectric performance.

## 4. Experimental Section

*Preparation of initial homogeneous amorphous ribbons*

Master alloy ingots were synthesized by melting a mixture of high-purity elemental constituents in a high-frequency induction furnace under an argon atmosphere. The resulting ingots were crushed into small fragments and loaded into a quartz tube equipped with a 5.0 mm × 0.8 mm nozzle. Amorphous ribbons were fabricated using a single-roll melt-spinning process, in which the ingot fragments were re-melted via induction heating and the molten alloy was ejected through the nozzle under an argon atmosphere at a pressure of 0.04 MPa onto a copper wheel rotating at 30 m/s. The nozzle-to-wheel gap was maintained at 0.2 mm throughout the process. The thickness of the ribbon was measured to be $20 \pm 2$ μm using a micrometer.

*Fabrication of heterogeneity graded amorphous ribbons*

Heterogeneity-graded amorphous ribbons were fabricated by annealing homogeneous amorphous ribbons under a temperature gradient naturally established within a horizontal tube furnace. The furnace consisted of a heating zone equipped with a resistive heating element and a cooling zone assisted by a fan (see **Fig. S1** for the setup). The ribbon was cut to a length of approximately 130 mm, with one end placed in the heating zone and the other in the cooling zone. This arrangement produced a temperature gradient along the ribbon length (i.e., *x*-direction) via heat conduction, leading to a gradual gradient of structural heterogeneity. Such a structure provides a suitable platform for investigating the influence of crystallinity and atomic-scale structural heterogeneity on transverse thermoelectric conversion.

Heat transfer and the resulting temperature gradient were assumed to be one-dimensional along the ribbon length, due to the large thermal mass of the surrounding quartz tube relative to the sample. The heating zone temperature was maintained at 773 K using the heating element, while the cooling zone temperature was kept near room temperature by a fan-assisted forced convection, with a wet cloth used to enhance thermal dissipation. The heating zone temperature increased at a ramp rate of 10 K/min, held at the set temperature for 15 min, and then cooled naturally. All heat treatments were performed under high vacuum ($<10^{-5}$ Pa) to prevent oxidation.

*Characterization of crystallinity using DSC*

Structural transitions associated with exothermic reactions were examined using DSC (Rigaku, Thermo plus EVO2). Measurements were carried out in a platinum crucible over a temperature range of 273–993 K at a heating rate of 10 K/min. Prior to the measurements, the samples were preheated at 373 K for 10 min to remove any residual moisture.

*Characterization of change in structural heterogeneity using XRD*

Structural characterization was carried out using a Rigaku MiniFlex600 X-ray diffractometer equipped with a Cr-Kα source ($\lambda = 0.22897$ nm). For consistency in comparison, the diffraction angles were converted to the commonly used Cu-Kα equivalent ($\lambda = 0.15406$ nm), as presented in **Figs. 2** and **S3**. The X-ray tube was operated at 40 kV and 15 mA. $\theta$-$2\theta$ scans were performed sequentially at 10 mm intervals along the $x$-position of the ribbon length.

*Measurement of position-dependent electrical conductivity*

The electrical conductivity ($\sigma$) of the sample was mapped using a standard four-probe method using a digital multimeter (Keithley, DMM6500). The four electrodes were connected using a probe scanning system equipped with needle probes. Specifically, two electrodes to apply a charge current were attached at both ends of the sample using silver epoxy in a line-contact configuration, ensuring uniform current flow along the sample length. Position-dependent voltage measurements were obtained using two point-contact probes, which were sequentially scanned along the sample. The sample was mounted on a rigid glass substrate, consistent with the setup used for XRD measurement.

*Measurement of position-dependent thermal diffusivity*

The thermal diffusivity ($D$) of the heterogeneity-graded FGM sample was measured using the spot periodic heating radiation thermometry method[59]. The sample was suspended between a modulated laser source (top) and an IR temperature detection system (bottom), with alignment maintained using a linear motor stage. The center of the suspended sample was periodically heated by a laser beam with a wavelength of 808 nm and a beam diameter of 150 µm. The laser was square-wave-modulated at a frequency of 5 Hz with a 50% duty cycle, resulting in an average power of 1.4 mW.

The resulting spatial heat diffusion was detected on the rear side of the specimen using the temperature detection system. Specifically, thermal radiation was collected using a Ge lens, and the temperature oscillations at each position were measured using an InSb IR detector cooled with liquid nitrogen. The phase lag between the modulated heating signal and the detected temperature response was analyzed using a lock-in amplifier. The values of $D$ were estimated using the phase lag in **Fig. 2b** and **Eq. (1)**.

*LIT-based AEE mapping*

The spatial distribution of AEE was characterized using a LIT system (DCG Systems Inc., ELITE). The samples were fixed onto a glass substrate using an electrically insulating adhesive. Measurements were conducted by applying a square-wave-modulated charge current with an amplitude of 500 mA under an in-plane magnetic field ($y$-direction) of $\pm 0.3$ T at room temperature ($T = 300$ K). The details of the measurements are described in Ref. [40,42,43].

*Structural analysis via APT and STEM*

Microstructural analysis was conducted using a FEI Titan G2 80-200 TEM. Elemental distribution was observed using APT in a laser mode with a CAMECA LEAP 5000 XS instrument, operated at a base temperature of 30 K and a laser pulse rate and energy of 250 KHz and 30 pJ, respectively. APT and STEM specimens were prepared using the lift-out technique with a FEI Helios 5UX dual beam-focused ion beam. APT data analysis was performed using CAMECA AP Suite 6.3 software.

*Measurement of anomalous Hall conductivity and Seebeck coefficient*

The anomalous Hall effect of the three representative samples was measured using a DC resistivity probe integrate in a He cryostat (CFMS, Cryogenics). Standard four-probe measurements were conducted with a current source (Keithley, 2450) and a nanovoltmeter Keithley, 2182a). A linear voltage-current response was confirmed by measuring 5 points with opposite current polarities. The Seebeck coefficient was measured at room temperature using a ZEM-3 system (ADVANCE RIKO).

**Author Contributions**

S.J.P. designed and conceived the study. K.U. supervised the project. R.G. and H.S.A. fabricated the initial as-spun amorphous metals, performed APT and TEM measurements, and analyzed the corresponding data. S.J.P. fabricated the heterogeneity-graded amorphous metals, measured and analyzed the XRD, DSC, Hall, Seebeck data, and performed the LIT measurement and analysis. S.J.P. and R.M. measured position-dependent electrical resistivity. S.J.P. measured and analyzed the position-dependent thermal diffusivity with help from Y.T. S.J.P. wrote the manuscript with contributions from all authors.

**Conflict of Interest**

The authors declare no conflict of interest.

**Acknowledgements**

The authors thank M. Goto for valuable discussion. This work was supported by ERATO "Magnetic Thermal Management Materials" (grant no. JPMJER2201) from JST, Japan.

**Data Availability Statement**

The data that support the findings of this study are available from the corresponding authors upon reasonable request.

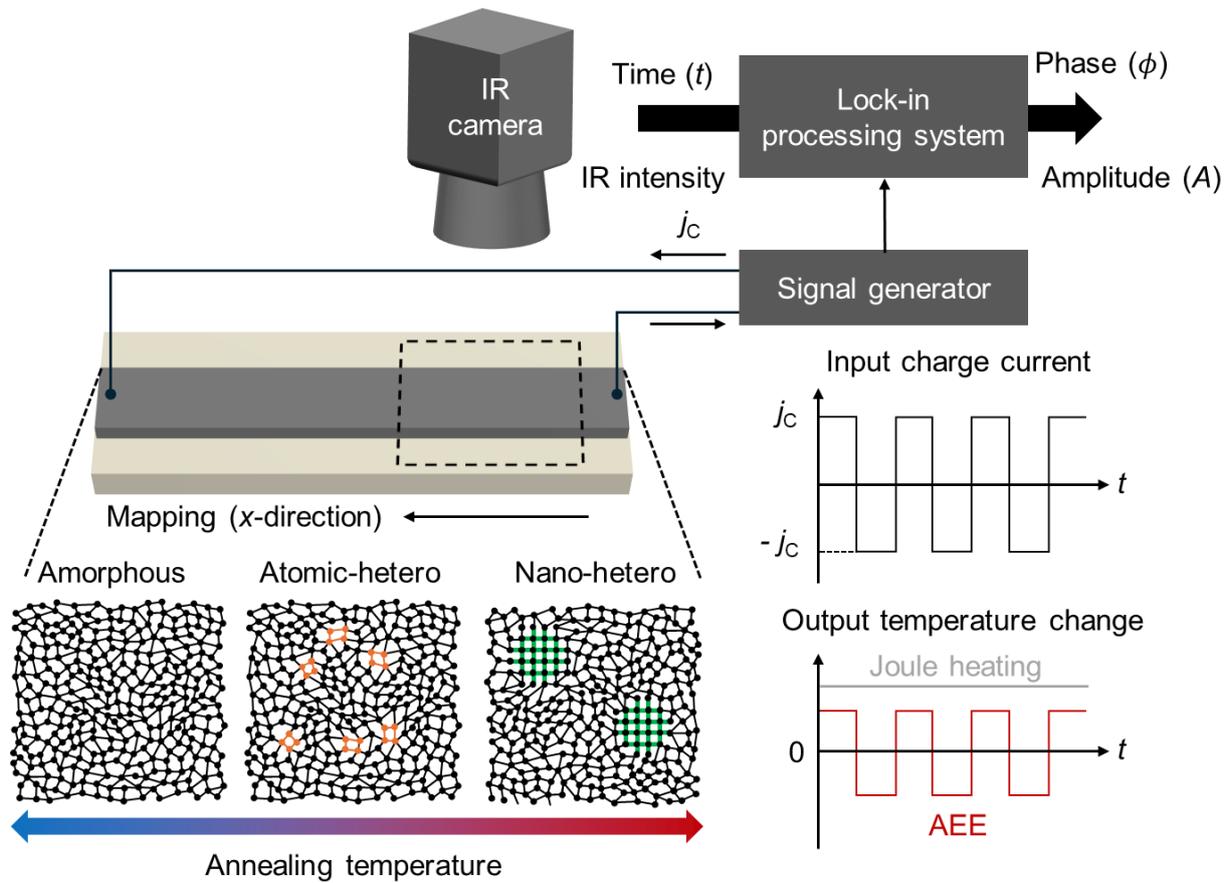

**Fig. 1 Schematic illustration of thermoelectric mapping of a functionally graded material (FGM) with controlled structural heterogeneity using lock-in thermography (LIT).** The FGM was fabricated by annealing under a temperature gradient, producing a one-dimensional heterogeneity-graded structure along the sample length ($x$-direction). LIT scans the sample using a square-wave-modulated input charge current $j_C$, enabling spatially resolved detection of the temperature modulation induced by the anomalous Ettingshausen effect (AEE).

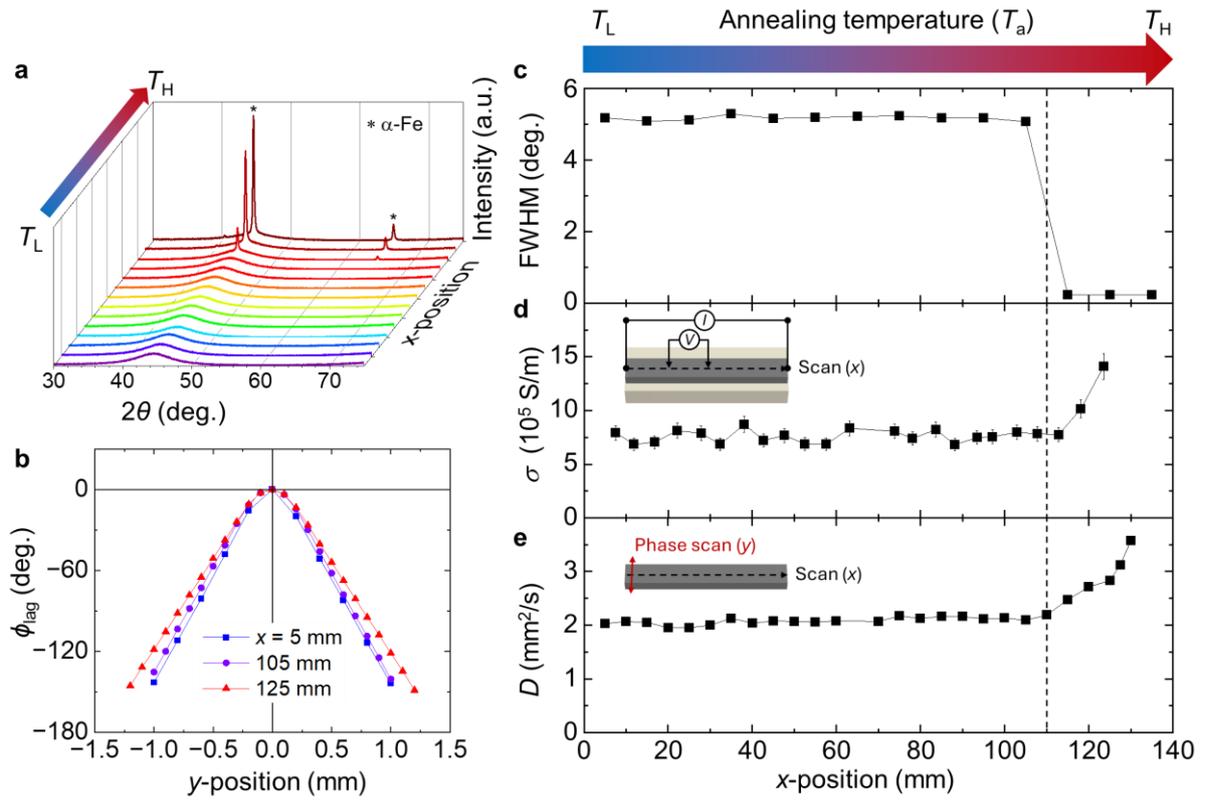

**Fig. 2 Characterization of structural and transport property variations in the heterogeneity-graded FGM ribbon along the length (*x*) direction.** (a) X-ray diffraction patterns for various *x*-positions. (b) Relative phase lag $\phi_{lag}$ as a function of *y*-position, referenced to the center position ($y = 0$). (c) Full width at half maximum (FWHM) of the main diffraction peak in (a). (d) Longitudinal electrical conductivity ($\sigma$), and (e) thermal diffusivity (*D*), both measured as a function of *x* with 5 mm intervals. The FWHM in (c) and *D* in (e) were extracted from the data in (a) and (b), respectively. Insets in (d) and (e) illustrate the measurement schematics for electrical and thermal transport characterization. The vertical dashed line in (c)–(e) indicates the crystallization onset.

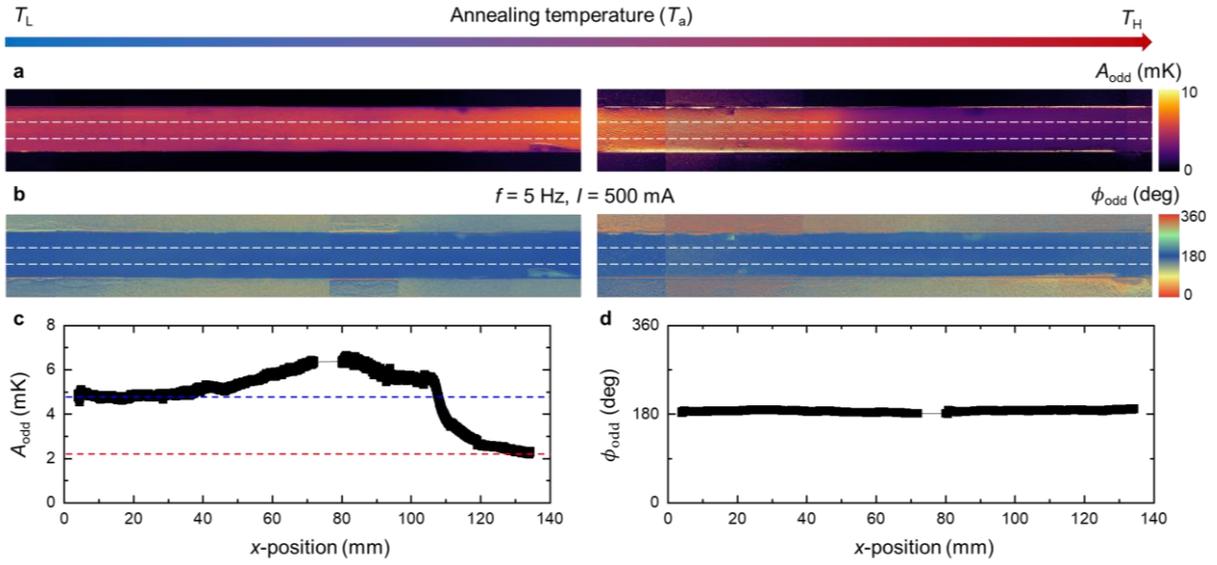

**Fig. 3 Thermoelectric mapping for a heterogeneity-graded amorphous alloy using lock-in thermography.** (a)–(b) Spatial maps of (a) the field-odd lock-in amplitude ($A_{\text{odd}}$) and (b) the corresponding field-odd phase ($\phi_{\text{odd}}$). (c)–(d) Averaged line profiles of (c) $A_{\text{odd}}$ and (d) $\phi_{\text{odd}}$ along $x$-direction, extracted from the highlighted regions in (a) and (b), respectively. The input current was modulated at a frequency $f = 5$ Hz and amplitude $I = 500$ mA. The horizontal dashed lines in (c) indicate the properties at $x = 5$ mm and 130 mm.

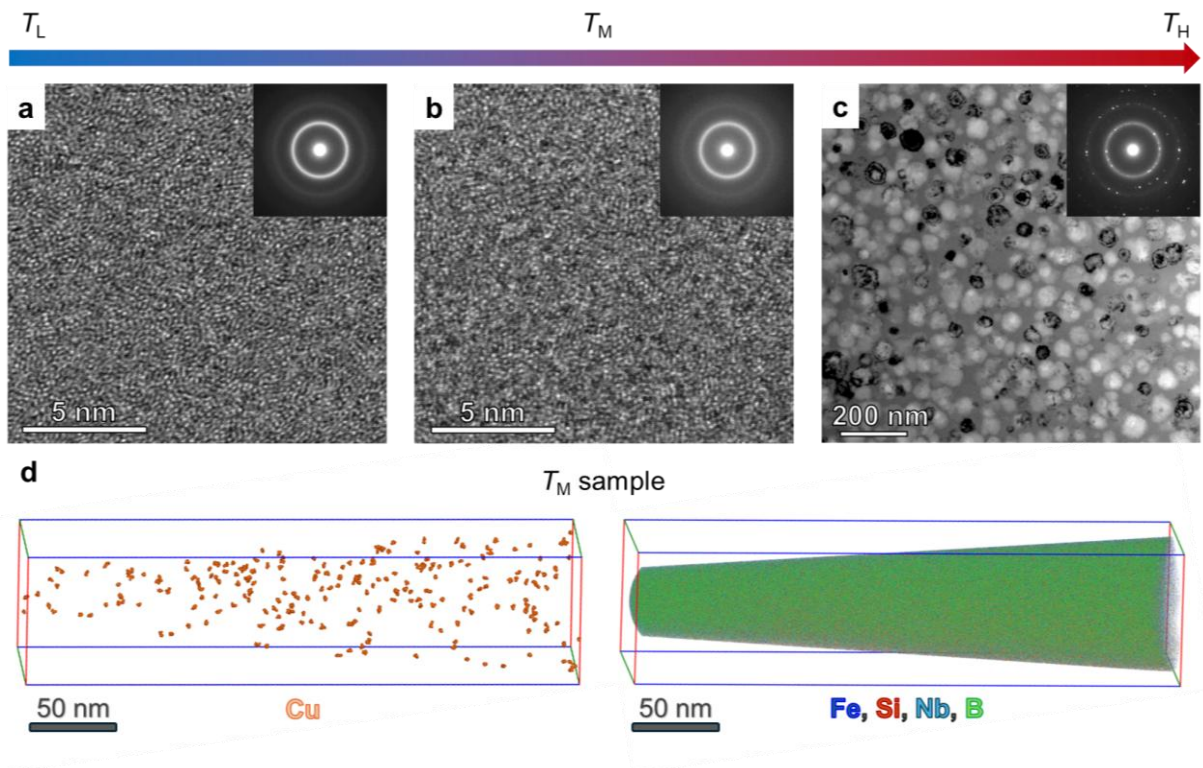

**Fig. 4 Structural characterization of heterogeneity-graded amorphous ribbons using** scanning transmission electron microscopy (STEM) and atomic probe tomography (APT). Bright field (BF)-STEM images taken from the samples in (a) low-temperature ($T_L$, $x$ = 5 mm), (b) mid-temperature ($T_M$, $x$ = 80 mm), and (c) high-temperature ($T_H$, $x$ = 125 mm) regimes. (d) Three-dimensional atomic distribution map of constituent elements obtained from $T_M$ sample by APT.

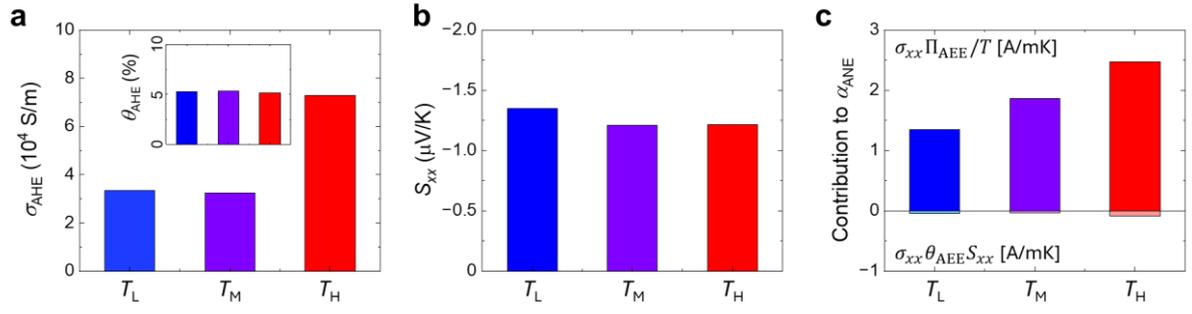

**Fig. 5 Transport properties of the three representative samples at $T_L$ ($x$ = 5 mm), $T_M$ ($x$ = 80 mm), and $T_H$ ($x$ = 125 mm).** (a) Anomalous Hall conductivity ($\sigma_{AHE}$) and anomalous Hall angle ($\theta_{AHE}$, inset), (b) Seebeck coefficient ($S_{xx}$), and (c) Decomposition of the anomalous Nernst conductivity $\alpha_{ANE}$ into the two contributing terms.

# Supplementary Information for

# Structural heterogeneity-induced enhancement of transverse magneto-thermoelectric conversion revealed by thermoelectric imaging in functionally graded materials

Sang J. Park[1,*], Ravi Gautam[1], Takashi Yagi[2], Rajkumar Modak[3], Hossein Sepehri-Amin[1] and Ken-ichi Uchida[1,3,*]

[1] National Institute for Materials Science, Tsukuba 305-0047, Japan

[2] National Institute of Advanced Industrial Science and Technology, Tsukuba 305-8563, Japan

[3] Department of Advanced Materials Science, Graduate School of Frontier Sciences, The University of Tokyo, Kashiwa 277-8561, Japan

*Correspondence to: PARK.SangJun@nims.go.jp (S.J.P.);
UCHIDA.Kenichi@nims.go.jp (K.U.)

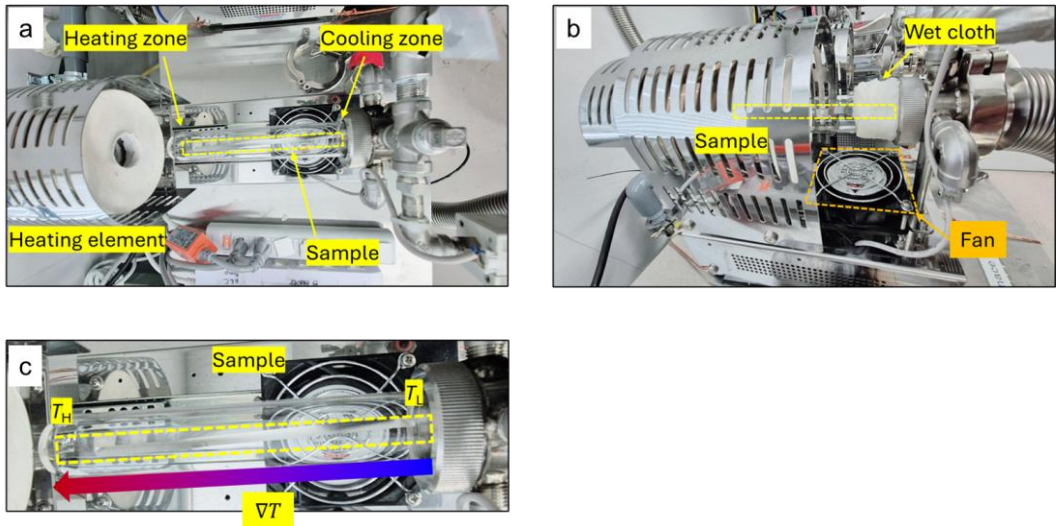

**Fig. S1 | Experimental setup for fabricating functionally graded materials with controlled disorder.** (a) Top view of the horizontal tube furnace. (b) Side view showing the cooling zone with a wet cloth to enhance cooling efficiency. (c) Top-view close-up of the furnace, illustrating the natural temperature gradient imposed on the sample

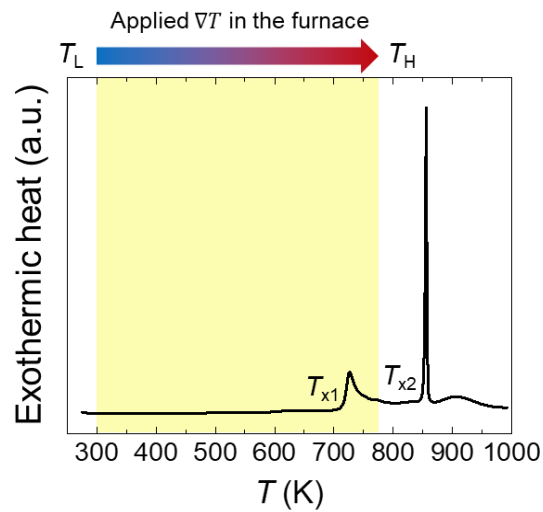

**Fig. S2 | Exothermic heat measured by differential scanning calorimetry of the sample.**

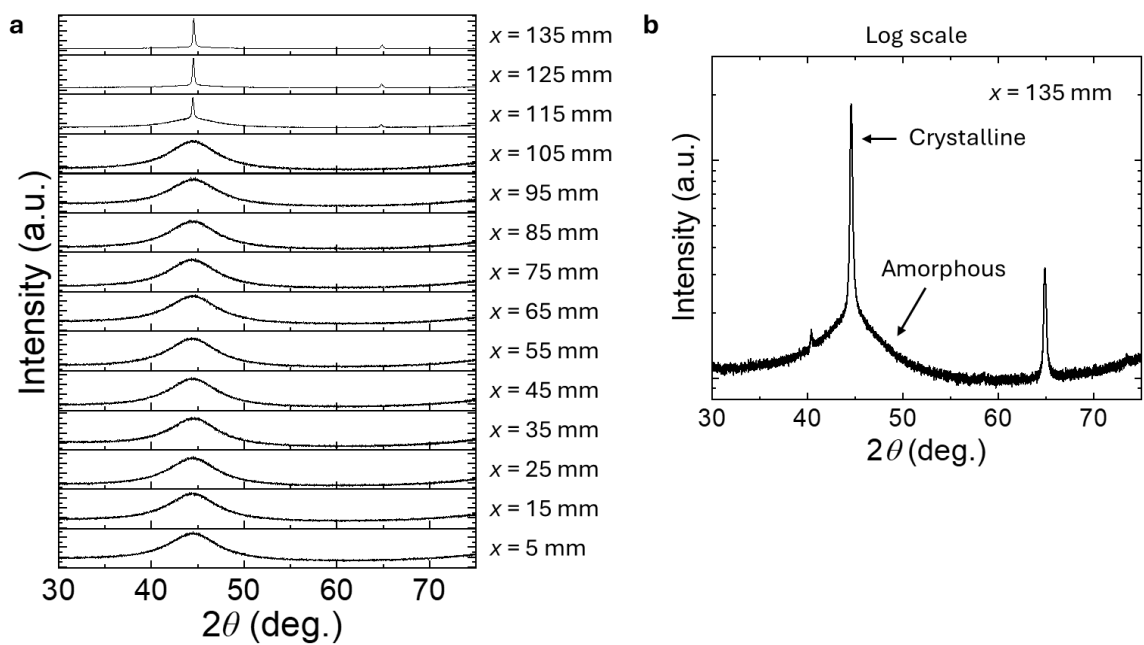

**Fig. S3 | Detailed XRD data.** (a) Position-dependent XRD patterns along the ribbon length (linear scale). (b) Log-scale XRD pattern at $x = 135$ mm.

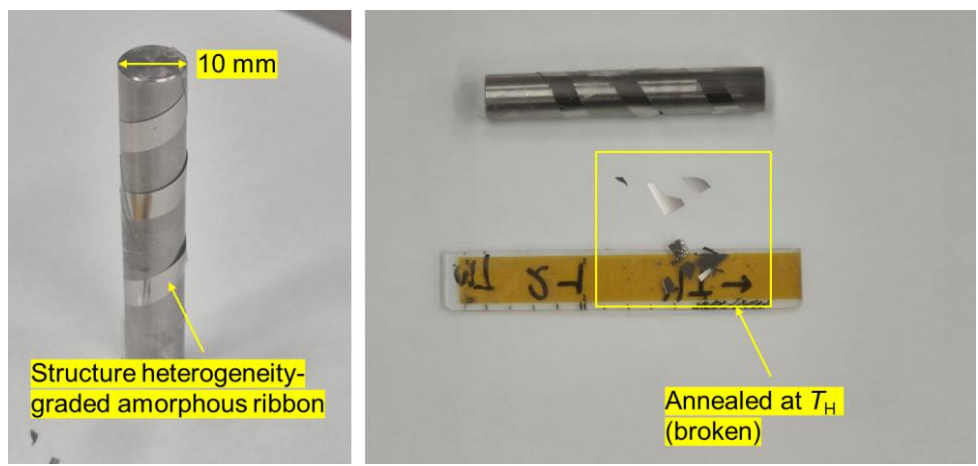

**Fig. S4 | Mechanical flexibility test.** The sample was attached to a round surface with a diameter of 10 mm. The overall sample retained mechanical flexibility; however, the crystalline regions annealed on the high temperature side ($T_H$) broke due to brittleness during handling, specifically when detaching the samples from the double-sided tape used for electrical conductivity measurements.